\documentclass[usenatbib, usegraphicx]{mn2e}

\newcommand{\aap}{\textit{A\&A}}

\newcommand{\apj}{\textit{ApJ}}
\newcommand{\aj}{\textit{AJ}}

\newcommand{\apjl}{\textit{ApJ Lett.}}
\newcommand{\apjs}{\textit{ApJ Suppl.}}

\newcommand{\mnras}{\textit{MNRAS}}
\newcommand{\nat}{\textit{Nature}}

\newcommand{\araa}{ARA\&A}

\newcommand{\msun}{M_\odot}
\def\aone{\mbox{$K_{a1}$}}
\def\atwo{\mbox{$K_{a2}$}}
\def\athree{\mbox{$K_{a3}$}}	
\def\ee{\mbox{e}}
\def\eg{e.g.,}
\def\dgone{\mbox{$D_{g1}$}}
\def\kgone{\mbox{$K_{g1}$}}
\def\kgtwo{\mbox{$K_{g2}$}}
\def\kgthree{\mbox{$K_{g3}$}}
\def\kgfour{\mbox{$K_{g4}$}}
\def\kgfive{\mbox{$K_{g5}$}}
\def\kgsix{\mbox{$K_{g6}$}}
\def\gyr{\mbox{Gyr}}

\def\kpc{\rm{kpc}}

\def\msun{\mbox{M$_\odot$}}

\def\nbody{\mbox{$N-$body}}

\def\omp{\mbox{$\Omega_p$}}

\def\pone{\mbox{$K_{p1}$}}
\def\ptwo{\mbox{$K_{p2}$}}

\def\sech{\mbox{sech}}

\def\mathnew{\mathsurround=0pt}
\def\simov#1#2{\lower .5pt\vbox{\baselineskip0pt
    \lineskip-.5pt\ialign{$\mathnew#1\hfil##\hfil$\crcr#2\crcr\sim\crcr}}}

\def\'#1{\ifx#1i{\accent"13\i}\else{\accent"13#1}\fi}

\title[Dynamics of barred galaxies]{Dynamics of barred galaxies: effects of disk height}

\author[Klypin et al.]{Anatoly Klypin$^1$,\thanks{e-mail: aklypin@nmsu.edu}
 Octavio Valenzuela$^2$, Pedro Col\'in$^3$, and Thomas Quinn$^4$\\
$^1$Department of Astronomy,  New Mexico State University, 
Las Cruces, NM 88003-0001\\
$^2$Instituto de Astronom\'ia,Universidad Nacional Auton\'oma de Mexico, A.P. 70-264,
04510, M\'exico, D.F.\\
$^3$Centro de Radiostronom\'ia y Astrof\'isica, UNAM, 
Apartado Postal 72-3 (Xangari), 58089 Morelia, Mexico\\
$^4$Department of Astronomy, University of Washington, Seattle, WA 98195-1580}
\date{Accepted 2009 Received August  2008}
\pagerange{\pageref{firstpage}--\pageref{lastpage}} \pubyear{2008}
\begin{document}
\label{firstpage}
\maketitle

\begin{abstract}
  We study dynamics of bars in models of disk galaxies embeded in
  realistic dark matter halos.  We find that disk thickness plays an
  important, if not dominant, role in the evolution and structure of
  the bars.  We also make extensive numerical tests of different
  $N$-body codes used to study bar dynamics. Models with thick disks
  typically used in this type of modeling (height-to-length ratio
  $h_z/R_d=0.2$) produce slowly rotating, and very long, bars. In
  contrast, more realistic thin disks with the same parameters as in
  our Galaxy ($h_z/R_d\approx 0.1$) produce bars with normal length
  $R_{\rm bar} \approx R_d$, which rotate quickly with the ratio of the
  corotation radius to the bar radius ${\cal R} =1.2-1.4$ compatible
  with observations. Bars in these models do not show a tendency to
  slow down, and may lose as little as 2-3 percent of their angular momentum
  due to dynamical friction with the dark matter over cosmological
  time. We attribute the differences between the models to a combined
  effect of high phase-space density and smaller Jeans mass in the
  thin disk models, which result in the formation of a dense central
  bulge. Special attention is paid to numerical effects such as the 
  accuracy of orbital integration, force and mass resolution.  Using
  three $N$-body codes -- Gadget, ART, and Pkdgrav -- we find that
  numerical effects are very important and, if not carefully treated,
  may produce incorrect and misleading results. Once the simulations
  are performed with sufficiently small time-steps and with adequate
  force and mass resolution, all the codes produce nearly the same
  results: we do not find any systematic deviations between the
  results obtained with TREE codes (Gadget and Pkdgrav) and with the
  Adaptive-Mesh-Refinement (ART) code.
 \end{abstract}

\section{Introduction}

Barred galaxies represent a large fraction ($\sim$65\% ) of all spiral
galaxies \citep[e.g.,][]{Eskridge00,Sheth08}. Bars are
ubiquitous. They are found in all types of spirals: in large
lenticular galaxies \citep{Aguerri05}, in normal spirals such as our
Galaxy \citep{Freudenreich98} and M31 \citep{Lia06,Beaton07}, and in
dwarf magellanic-type galaxies \citep{Valenzuela07}.  An isolated
stellar disk embeded into a dark matter halo spontaneously forms a
stellar bar as a result of the development of  global disk instabilities
\citep[e.g.,][Sec. 6.5]{BT}. Bars continue to be closely
scrutinized because of their connection with the dark matter halo
\citep[\eg][]{od03, hbk05, cvk06, lia07}.  Because the bars rotate
inside massive dark matter halos, they lose some fraction of the
angular momentum to their halos and tend to slow down with time
\citep{Tremaine84,Weinberg85}.

The formation of bars and associated pseudobulges is often considered
as an {\it alternative} to the hierarchical clustering model
\citep{Kormendy04}. This appears to be incorrect: recent cosmological
simulations indicate that the secular bulge formation is a {\it part
  (not an alternative) of the hierarchical scenario}. The simulations
of the formation of galaxies in the framework of the standard
hierarchical cosmological model indicate that bars form routinely in
the course of assembly of halos and galaxies inside them
\citep{Mayer08,Ceverino08}.  The simulations have a fine resolution of
$\sim 100$\,pc and include realistic treatment of gas and stellar
feedback, which is important for the survival of a bar. Bars form
relatively late: well after the last major merger ($z\approx 1-2$ for
normal spiral such as our Milky Way), when a collision of gas rich
galaxies brings lots of gas with substantial angular momentum to the
central disk galaxy. As the disk accretes the cold gas from the halo,
it forms a new generation of stars and gets more massive. At some
stage, the disk becomes massive enough to become unstable to bar
formation.  Once the stellar bar forms, it exists for the rest of the
age of the Universe.

The cosmological simulations are still in preliminary stages, and it
is likely that many results will change as they become more accurate
and the treatment of the stellar feedback improves.  However, existing
cosmological simulations already show us the place and the role of
traditional $N$-body simulations of barred galaxies, which start with
an unstable stellar disk. It was not clear whether and how this
happens in the real Universe. Now the cosmological hydrodynamic
simulations tell us that this is somewhat idealistic, but still a
reasonable setup compatible with cosmological models.

Simulations of bars play an important role for understanding the
phenomenon of barred galaxies
\citep[e.g.,][]{Miller78,Sellwood80,lia03,vk03}. Numerical models
successfully account for many observed features of real barred
galaxies \citep{Bureau05,Bureau06,Beaton07}. So, it is very important
to assess the accuracy of those simulations.  Recent disagreements
between results of different research groups \citep{vk03,od03,sd06}
prompted us to undertake a careful testing of numerical effects and to
compare results obtained with different codes. This type of code
testing is routine in cosmological simulations
\citep{Frenk99,Heitmann05,Heitmann07}, but it never has been done
before for bar dynamics. Testing and comparison of numerical codes is
important for validating different numerical models. It was
instrumental for development of precision cosmology. 
It is our goal to make such tests for $N$-body models of barred galaxies.

We use four different popular $N$-body codes: ART \citep{KKK97},
Gadget-1 and Gadget-2 \citep{volker01,volker05}, and Pkdgrav
\citep{wsq04}.  ART is an Adaptive Mesh Refinement code that reaches
high resolution by creating small cubic cells in areas of high
density.  Gadget and Pkdgrav, on the other hand, are TREE codes that
compute forces directly for nearby particles and use a multipole
expansion for distant ones. We use the codes to run a series of
simulations using the {\it same initial conditions} for all codes.

We also address another issue: the effects of disk thickness on the 
structure and evolution of bars. Only recently have the simulations started
to have enough mass and force resolution  to resolve the
vertical height of stellar disks. We use different codes to show that
the disk height plays an important and somewhat  unexpected role.

One of the contentious issues in the simulations of bar dynamics is the angular
speed and the structure of bars in massive dark matter halos.  The
amount and the rate at which bars slow down is still under
debate. \citet{ds98,ds00} find in their massive halo models, i.e.,
those for which the contributions of the disk and the halo to the mass
in the central region are comparable, that the bar loses about 40\%
of its initial angular momentum, $L_z$, in $\sim 10\ \gyr$. However,
in simulations with much better force resolution and a more realistic
cosmological halo setup, \citet{vk03}  and  \citet{cvk06} find a decrease in $L_z$ of
only 4--8\% in $\sim 6\ \gyr$.  \citet{ds98,ds00} also find that bars
do not significantly slow down for lower density halos.

\citet{vk03} presented bar models in which stellar disks were embedded
in a CDM Milky-Way-type halos with realistic halo concentrations $c
\sim 15$, where $c$ is the ratio of the virial radius to the
characteristic radius of the dark matter halo. These simulations were
run with the ART code  with high spatial resolution of 20-40
pc. The bars in the models were rotating fast for billions of years.
\citet{vk03} argued that slow bars in previous simulations were an
artifact of low resolution. \citet{sd06} used initial conditions of one of the models of
\citet{vk03} and run a series of simulations using their hybrid,
polar-grid code.  They found, in most cases, a different evolution 
than that reported in Valenzuela \& Klypin. In particular,
contrary to Valenzuela \& Klypin's results, they did not find that the
bar pattern speed is almost constant for a long period of time.  They
attribute the differences to the ART refinement scheme.  

While \citet{vk03} mention numerical effects (lack of force and mass
resolution) as the cause for excessive slowing down of bars in earlier
simulations, there was another effect, which was not noticed by
\citet{vk03}: Their disk was rather thin, with a scale-height $h_z$ of
only 0.07 of the disk scale-length $R_d$. This should be compared with
$h_z \sim 0.2\ R_d$ used in most studies of stellar bars
\citep[e.g.,][]{lia02,lia03,m-v06}.  Models of \citet{ds98} have $h_z
= 0.1 R_d$, but the resolution of their simulations was grossly
insufficient to resolve this scale. \citet{od03} used $h_z = 0.1 R_d$
for a model, which had very little dark matter in the central disk
region: $M_{\rm dm}/M_{\rm disk} \approx 1/4$ inside radius $R=3R_d$.
Dependence of bar speed on disk thickness was noted by \citet{lia00}:
thicker disks result in slower bars.

The remainder of this paper is organized as follows. In
section~\ref{sec:height} we give a detailed review of available data
on disk scale heights and present a simple analytical model for the
relation of the disk scale length with the disk scale height. We
describe our numerical models in Section~\ref{sec:Models}. In
Section~\ref{sec:Numerics} we give a brief description of the codes
and present analysis of numerical effects.  Main results are presented
in Section~\ref{sec:Results}. We summarize our results in
Section~\ref{sec:Discussion}.

\section{Disk heights}
\label{sec:height}
Disk scale height appears to play a very important role in the
development of barred galaxies. Thus, it is important to know 
the range of disk scale heights in real spiral galaxies.  The most
accurate measurement of the disk height  comes from the Milky Way
galaxy. Edge-on galaxies provide another opportunity to measure disk
heights, but those measurements are much less accurate because of dust
absorption close to the disk plane.  There is not much disagreement
between different studies regarding the disk thickness of the Milky
Way: the exponential scale height of the stellar thin disk is $h_z
\approx 300$\,pc \citep{Gilmore83,Ojha99,Juric08}. The thick disk
component has a scale height of $\sim 1$~kpc, but it has a small
fraction of mass. Neutral hydrogen and molecular gas have scales
200~pc and 50~pc.  So, we can estimate the scale height of the mass
distribution as 250-300~pc. Using the exponential disk scale length
$R_d\approx 3$\, kpc \citep[e.g,][]{Dehnen98,Klypin02}, we get the
ratio of the scale height to the scale length $h_z/R_d \approx 0.1$.

Observations of edge-on galaxies can be used to estimate the disk
heights for other galaxies \citep[e.g.,][]{Kruit81,Kregel02,
  Bizyaev02,Kregel05,Yoachim06}.  \citet{Kregel05} gave ratios of
exponential heights to exponential lengths for 34 edge-on galaxies
measured in  $I$-band and found that the median ratio is $h_z/R_d
\approx 0.12$. \citet{Seth05} present fits for brightness profiles in
$K$-band (2MASS images) for two Milky Way-type galaxies $NGC891$
($V_{\rm max}=214$~km/s) and $NGC4565$ ($V_{\rm
  max}=227$~km/s). They find the half-light height to
exponential disk scale ratios $z_{1/2}/R_d=0.072$ and 0.085 for the two
galaxies correspondingly. For the five galaxies in the
\citet{Yoachim06} sample, which had circular velocities in the range
$150-200$~km/s, the average ratio was $z_{1/2}/R_d\approx 0.1$.

Estimates of disk heights in edge-on galaxies suffer from substantial
absorption close to the plane of the disk
\citep{Xilouris99,Yoachim06}. This makes scale heights of the thin disk
difficult to measure directly and causes the results to be
dominated by flux coming from high galactic latitudes, where the thick
disk is dominant. In turn, this leads to a substantial overestimation
(by a factor of 2--3) of the disk heights even in red bands, if one
interprets those as estimates of the thin disk component
\citep{Yoachim06}.

We can use stellar-dynamical arguments to estimate the
scale-heights. The idea is to use the ratio of the vertical velocity
dispersion $\sigma_z$ to the radial velocity dispersion
$\sigma_R$ \citep{Kregel05}. For old main sequence stars ($B-V>0.5$) in the solar
neighborhood, the ratio is measured to be $\sigma_z/\sigma_R=0.5-0.6$
\citep{Dehnen98b}. We find the same ratio for most of our dynamical
model. The vertical velocity dispersion is related to the disk 
scale-height, and the radial velocity dispersion is related to the disk
scale-length (among other parameters of the disk). 

Assuming an exponential stellar disk with the vertical density profile
$\sech^2(z/h_z)$, one gets: 
\begin{equation}
\sigma_R(R)= Q \frac{3.36 G \Sigma(R)}{\kappa(R)}, \hspace{0.5cm}
\sigma_z^2 (R) = \pi G h_z \Sigma(R),
\label{eq:sigmas}
\end{equation}
where $Q$ is the Toomre stability
parameter; $\Sigma(R)$ is the surface density, and $\kappa(R)$ is the
epicycle frequency. For galaxies with flat rotation curves and for
radii $R>R_d$ we can use $\kappa =\sqrt{2}V_{\rm circ}/R$. The
circular velocity $V_{\rm circ}$ is defined by the mass distribution
given by the sum of three components: disk, bulge, and dark matter. It
is convenient to parameterize those relative to the total disk mass:
$M(R)/M_{\rm disk} =f_{\rm disk}(R)+M_{\rm bulge}/M_{\rm disk} +M_{\rm
  dm}(R)/M_{\rm disk}$, where $f_{\rm disk}(R)$ is the fraction of the
disk mass inside radius $R$.  Combining these relations, we get the
following expression for the height-to-length ratio:
\begin{equation}
\frac{h_z}{R_d}=\left(\frac{\sigma_z}{\sigma_R} \frac{3.36Q}{2\pi}  \right)^2
\frac{X^3\exp(-X)}{f_{\rm disk}(X)+f_{\rm bulge}+f_{\rm dm}(X)},
\label{eq:height}
\end{equation}
where $X=R/R_d$,
$f_{\rm bulge}=M_{\rm bulge}/M_{\rm disk}$, and $f_{\rm dm}(X)=M_{\rm dm}(X)/M_{\rm disk}$,

We can now estimate the disk height at different radii. For example,
we can get it at $R=3R_d$ assuming that the mass of dark matter is
about equal to the disk mass $f_{\rm dm}=1$ \citep{Klypin02,Widrow08} and
taking a small bulge $f_{\rm bulge}=0.2$. For $Q=1.5$  \citep{Widrow08} and taking
$\sigma_z/\sigma_R=0.5$ we get $h_z=0.11R_d$, which is consistent with
the height of Milky Way disk. Eq.~(\ref{eq:height}) gives  nearly the same
height for  $R_d<R<3R_d$, if we scale the dark matter contribution in
such a way that the rotation curve stays constant. 

To summarize, the disk scale-heights are relatively small for high
surface brightness galaxies such as our Milky Way with 
$h_z=0.1R_d$ being a reasonably accurate estimate. Care should be
taken not to over interpret results from edge-on galaxies.

\section{Models and simulations}
\label{sec:Models}
\subsection{Initial conditions}

The setup of  initial conditions is described in detail in \citet{vk03}. 
Here we briefly summarize the most important features. The system of a halo
and a disk, with no initial bulge or  bar, is generated using the
method of \cite{hern93}. In cylindrical coordinates the density of the
stellar disk is approximated by the following expression:
\begin{equation}
\rho_d (R,z) = \frac{M_d}{4\pi h_z R_d^2} \ee^{-R/R_d} \sech^2\left(\frac{z}{h_z}\right),
\end{equation}
where $M_d$ is the mass of the disk, $R_d$ is the scale length, and
$h_z$ is the disk scale height. The latter is assumed to be constant
through the disk. The radial and vertical velocity dispersion are
given by eqs.~(\ref{eq:sigmas}).  Our models keep $Q$ fixed along the
disk.  The azimuthal velocity and its dispersion are found using the
asymmetric drift and the epicycle approximations.

The models assume a NFW density 
profile \citep{nfw97} for the dark matter halo component,
which is described by 
\begin{eqnarray}
\rho_{DM}(r) = \frac{\rho_0}{x(1+x)^2}, \hspace{0.5cm} x \equiv r/r_s, \\
M_{vir} = 4\pi \rho_0 r_s^3 \left[ \ln(1+c) - \frac{c}{1+c} \right],  \hspace{0.2cm} c =  
\frac{R_{vir}}{r_s},
\end{eqnarray}
where $M_{vir}$, $R_{vir}$, and $c$ are the virial mass, the virial radius, 
and concentration of the halo, respectively. Given $M_{vir}$, the virial 
radius is found once a cosmology is adopted\footnote{We adopt the 
flat cosmological model with a non-vanishing cosmological constant 
with $\Omega_0=0.3$ and $h=0.7$.}. Equations~(4-56) of \citet{BT}
and the assumption of isotropy in the velocities allow us 
to determine the radial velocity dispersion as
\begin{equation}
\sigma^2_{r,DM} = \frac{1}{\rho_{DM}} \int_r^\infty \rho_{DM} \frac{GM(r)}{r^2} dr,
\end{equation}
where $M(r)$ is the mass contained within radius $r$ and $G$
the gravitational constant. 

\begin{table*}
\caption{Initial Parameters of the models}
\begin{tabular}{lccccccc}
\hline
Code & Name & N$_{disk}$ & N$_{total}$ & 
N$_{eff}$ & Force resolution & Time-step & Disk scale  height $h_{z}$ \\
 &  & ($10^5$) & ($10^6$) & ($10^6$) & (pc) & ($10^4$yr) & (pc)  \\
(1) & (2) & (3) & (4) & (5) & (6) & (7) & (8)\\
\hline
 &\multicolumn{7}{l}{K series models: M$_{disk}= 5 \times 10^{10} \msun$ 
M$_{tot}= 1.43 \times 10^{12} \msun$ R$_d = 3.86\ \kpc$ $Q = 1.8$ $c = 10$}\\ 
    ART        &  \aone   &  2.33  & 2.7   &  6.6  &  44  &  1.4  & 200     \\ 
     ART      &  \atwo    &  2.33  & 2.3   &  6.6  &  86   &  1.9  &  200    \\ 
   ART            & \athree   & 2.00  & 2.2  &  5.9  &  170 &   2.2  & 714   \\ 
Gadget-1  & \kgone    & 1.00  & 1.1  &  2.9  &  280 &   8.6  & 714   \\ 
 Gadget-2 & \kgtwo   &  2.33  & 2.5   &  6.7  & 112  &  2.6  & 200   \\ 
 Gadget-2  & \kgthree & 4.67  & 5.0  & 13.8 &  112 &  3.3  & 200   \\ 
 Gadget-2  & \kgfour & 2.00    & 2.2  &  5.9  &  140  &  1.4  & 200    \\ 
Gadget-2  & \kgfive     & 1.00  & 1.1  &  2.9  &  280 &  29.2  & 714   \\ 
Gadget-2  & \kgsix     & 1.00  & 1.1  &  2.9  &  280 &  3.6  & 714   \\ 
Pkdgrav     & \pone      & 1.00  & 1.1  &  2.9  & 136  &  24.5  & 714    \\ 
Pkdgrav     & \ptwo      & 2.33  & 2.5  &  6.7  & 136  &  1.2  & 200    \\ \\
 &\multicolumn{7}{l}{ Model D: M$_{disk}= 5 \times 10^{10} \msun$ 
M$_{tot}= 1.43 \times 10^{12} \msun$ R$_d = 2.57\ \kpc$  $Q=1.3$ $c=17$} \\ 
Gadget-2   & \dgone   & 2.33  & 2.5  &   6.7 & 112 & 2.6  & 200 \\ 
\hline
\end{tabular}
\end{table*}

\subsection{Description of the models}

Selection of parameters of our models is motivated by a number of
reasons.  First, to simplify the comparison with previous results, we
chose parameters, which are close to those used in
\citet{cvk06}. Indeed, some of our models have exactly the same
parameter as models $K_{hb}$ and $D_{cs}$ in \citet{cvk06}. In this paper we
preserve the first letter of the model name (K or D), but use
subscripts to identify numerical code. Second, in order to test the
effects of the disk height, we construct a new model by taking the
$K_{hb}$ model and giving it a larger scale-height $h_z=714$~pc
instead of $h_z=200$~pc. Third, our models are motivated by
predictions of cosmological models. Thus, the models have extended
dark matter halos with the NFW profile, realistic virial masses, and
concentrations. The central regions of the models are dominated by the
disk.  The radius at which initially the dark matter mass is equal to
the disk mass is equal to 9~kpc for models K and to 6~kpc for models
D. Initial profiles of different components are presented in the top
two panels of Figure~1 in \citet{cvk06}.

The disk scale-height increases in the course of evolution. We find that for
thin disk models it more than doubles after 5~Gyrs of evolution, while
for disk thick models the scale-height increases less: by a factor 
of 1.6. As a result, the scale-height to scale-length ratio of 
evolved thin disk models is close to the observed $h_z/R_d=0.1$.
The ratio for evolved thick disk models is, on the other hand, twice 
the observed one.

Table~1 presents the parameters of the models. The first and the
second columns give the name of the code and the name of the
model. The capital letter of model name represents the model type (K
or D). The first subscript in the model name indicates the code used
to make the run: {\it a} for ART, {\it g} for Gadget, and {\it p} for
Pkdgav.  In columns (3), (4), and (5) we show the disk, the total
(disk + dark matter), and the effective number of particles -- the
number of particles, which we would need, if we used equal-mass
particles.  The force resolution (6-th column) is twice the smallest
cell size for ART code and the spline softening (2.8 times the
effective Plummer softening) for tree codes. Note that the force
resolution is the distance at which the force accurately matches the
Newtonian force. The force continues to increase even below the
resolution resulting in large changes in density. Column~7 shows the
smallest time-step of simulations. All codes use variable time-steps. Details are
given in the next section.

The number of particles inside a sphere of radius equal to the force
resolution is quite substantial. For a typical simulation such as
\atwo~ or \kgtwo~ with the resolution $\sim 100$~pc, there are $\sim 100$
particles inside the resolution radius at the center of the system at
the initial moment.  For TREE codes, which keep the resolution
constant, the number declines with distance, but it is still
large in the plane of the disk. For example, it is $\sim 10-15$ at
8~kpc. The ART code maintains the nearly constant number of particles
inside (increasing) radius of the force resolution at the level of
60~particles \citep[see details in][]{cvk06}.

Particles with different masses are used in our simulations to
increase the mass and force resolution in the central disk region. This is done  by
placing many small-mass particles in the central disk-dominated region
and by utilizing large-mass particles in the outer halo-dominated
areas.  We use four mass species. The first species represents disk
and dark matter particles in the central halo region (the central
$\sim 40$~kpc region). Both the disk particles and the central dark
matter particles have the same mass. More massive particles rarely
enter the central $\sim 10$~kpc region.  The mass species differ by a
factor of two between one species and the next.

The mass resolution -- the mass of a disk particle or the mass of the
smallest dark matter particle -- is given by the ratio of the disk
mass to the number of disk particles. It is in the range $m_1 =(1 -
5)\times 10^5\msun$.  We present the time-step and the initial scale
height in the last two columns. The typical number of time-steps for
simulations is $(2-4)\times 10^5$.  The physical parameters of the
models, such as the mass of the disk or concentration of the dark
matter halo, are shown in separate rows.

In order to estimate the bar pattern speed \omp\ we first determine
the orientation of the bar by iteratively applying the method of the
inertia tensor in the plane of the disk. \omp\ is obtained
subsequently by numerical differentiation: $\omp = d\phi/dt$, where
$\phi$ is the position angle of the bar. In practice, we use about ten
consecutive snapshots for which the increasing function $\phi$ is
available and make a least squares fit. Then \omp\ is given by the
slope of the straight line. The bar amplitude $A_2$ is computed
similar to \cite{vk03}. For each logarithmically spaced cylindrical
bin we find the amplitude of the second Fourier harmonic. Then the
amplitude is smoothed over the radius and the maximum is taken as the
bar amplitude $A_2$. The bar length is defined as out-most radius at
which the ratio of axes of isosurface density contours is 1.5. When
finding radius of corotation, we use rotational velocity curve of disk
particles $V_{\rm rot}(r)$ and find radius, at which it is equal to
$\omp r$.

\section{Numerical effects}
\label{sec:Numerics}
\subsection{Codes}

\begin{figure}
\includegraphics[width=0.45\textwidth]{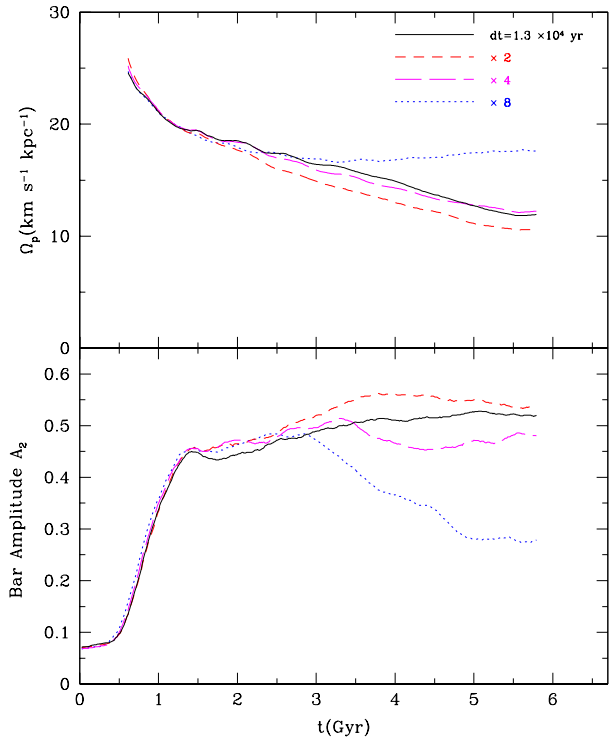}
\caption[fig:DTcon]{
Effects of the time-step on the evolution of bars. The time-step for
each model is indicated in the right corner of the top panel.
Evolution of the pattern speed of the bar (top panel) and the bar
amplitude (lower panel) is shown for a thin disk K model with force
resolution 560~pc.  Models only start to deviate from each other at
the beginning of the buckling phase ($t\approx 2.5-3$~Gyrs), which is
marked by a slight drop in the amplitude of the bar. Once this stage
is passed models with sufficiently small time-steps start to converge
again except for the run with the largest time-step. We find that the
time-step $dt\approx 10^5$~yrs is not sufficiently small for this
model and produces incorrect results.
\label{fig:DTcon}}
\end{figure}

Simulations were run with three $N$-body codes: ART \citep[Adaptive
Refinement Tree,][]{KKK97}, Gadget-1 or Gadget-2
\citep{volker01,volker05}, and Pkdgrav \citep{wsq04}. ART is an
adaptive mesh refinement (AMR) $N$-body code, which achieves high
spatial resolution by refining the base uniform grid in all
high-density regions with an automated refinement algorithm.  Gadget
is a parallel TREE code. Here we only use the $N$-body part of the
code. Pkdgrav is another TREE code. These two TREE codes differ in the
characteristics of the gravitational tree algorithm. For instance,
Gadget employs the Barnes \& Hut TREE construction \citep{bh1986}
while Pkdgrav uses a binary K-D TREE \citep{bentley1979}. Gadget-2
only uses monopole moments in the multipole expansion while Pkdgrav
advocates a hexadecapole as the optimum choice. (Gadget-1 can be
configured to use octopole moments). The codes also differ in the
cell-opening criterion.  Here we use an opening angle criterion of
$\theta = 0.7$ for Pkdgrav.  In the case of Gadget-2 runs we use a
tolerance parameter $\alpha =0.005$ (see eq.(18) in \citet{volker05}).
The main advantage of Gadget-2 as compared with Gadget-1 is a more
accurate time-stepping scheme \citep{volker05}.

The codes use different variants of 
leapfrog scheme. An algorithm of integration of trajectories can be written
as a sequence of operators, which advance particle positions (called
drifts) and changes velocities (called kicks). For example, a simple
constant-step leapfrog scheme is:
\begin{eqnarray}
v(t_{n+1/2}) = v(t_{n-1/2})+g(t_n)dt \quad {\rm Kick}, \\
x(t_{n+1}) = x(t_{n})+v(t_{n+1/2})dt \quad {\rm Drift}. 
\end{eqnarray}
Thus, the leapfrog integration is a sequence of
$K(dt)D(dt)K(dt)D(dt)$... operators.  We call this a KD scheme. Note
that the order of operators is not important: the DK scheme is
identical to the KD scheme. When the time-step changes with time, the
order of operators makes a difference. It is also important to select
the moment, at which the time-step should be changed. Following
\citet{Quinn97} we use the operator $S$ to indicate the moment when a new
time-step is selected. We also need to specify the time when
accelerations (in kicks) and velocities (in drifts) are estimated
relative to the moment to which the velocities and coordinates are
advanced. We attach the sign '+' ('-') to the name of operator to
indicate that the operator uses information from the beginning (end)
of the time-step. For example, $K_-(dt)$ is:
$v(t_n)=v(t_{n-1})+g(t_{n-1})dt$ and the operator $D_+(dt)$ is
$x(t_n)=x(t_{n-1})+v(t_{n})dt$.  Operators with subscript 0 are time
symmetric: they use information at the middle of the time-step.  Using
these operators we can write the algorithms of time-stepping in all
our $N$-body codes:
\begin{eqnarray}
SD_-(dt/2)K_0(dt)D_+(dt/2), \quad {\rm Gadget-1}\\
SK_-(dt/2)D_0(dt)K_+(dt/2), \quad {\rm Gadget-2}\\
K_+(dt_1/2)SK_-(dt/2)D_0(dt), \quad  \quad  \ \  \quad {\rm ART}\\
D_-(dt/2)SK_0(dt)D_+(dt/2). \quad   {\rm Quinn\ et\ al.} 
\label{eq:schemes}
\end{eqnarray}

\begin{figure}
\includegraphics[width=0.45\textwidth]{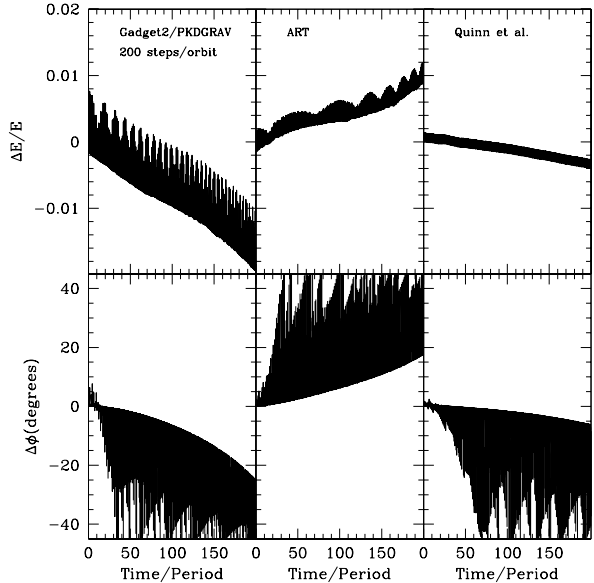}
\caption{
Errors of orbit integration. We integrate trajectory of a particle
moving in a gravitational potential created by isothermal distribution
of matter $\phi=\log(r)$, $\rho\propto r^{-2}$. The particle has
angular momentum 1/5 of the circular orbit and moves on an elongated
orbit with $r_{\rm max}/r_{\rm min}\approx 10$, which is not unusual
for orbits in strong bars. The number of time-steps is 200 per radial
period, which corresponds to the time-step of $10^5$~yrs when scaled to a realistic
galaxy model. The top panels show the error in energy conservation and the
bottom ones are the errors in position angle. The left column is for a
the leap-frog scheme implemented in Gadget-2 and Pkdgrav codes. The middle (right) column is for
 ART (Quinn et al.) code. The energy
conservation is reasonably small, but errors in the orbital angle are
unacceptably large.  
\label{fig:Move1}}
\end{figure}

The Pkdgrav code has different integration schemes. The scheme given in
eqs.~(\ref{eq:schemes}) uses the algorithm described by
\citet{Quinn97}. In our simulations with the Pkdgrav code, we use the
scheme which is identical to the Gadget-2 code both in the sense of
the  sequence of stepping and refinement conditions.  This is, in
fact, the way Pkdgrav is most commonly run \citep{wsq04}.  The ART code
uses the time-step $dt_1$ from previous moment to start the
integration. Then, having information on coordinates, it makes a
decision on the value of new time-step. 
The Gadget-2 and the ART schemes look different, but
actually they are identical, which can be seen   when one writes the
sequence of a few time-steps. The Gadget-1 and the Quinn et al.  schemes look
very similar, but they are quite different: the position of the $S$
operator makes the Quinn et al. scheme more accurate \citep{Quinn97}.

\begin{figure}
\includegraphics[width=0.45\textwidth]{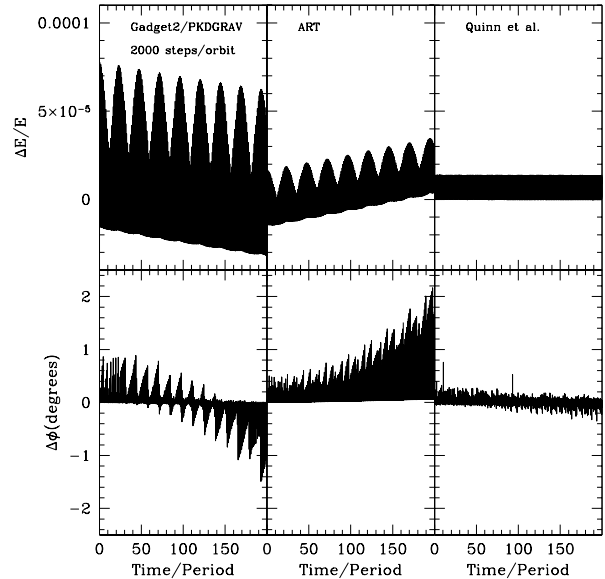}
\caption{
Errors of orbit integration. The same as in Figure~\ref{fig:Move1},
but for 2000 time-steps per radial period. This corresponds to a
time-step $\sim 10^4$~yrs in realistic simulations.  The errors of
integration are very small for both the energy conservation and for
the orbital angle.
\label{fig:Move2}}
\end{figure}

\begin{figure}
\includegraphics[width=0.45\textwidth]{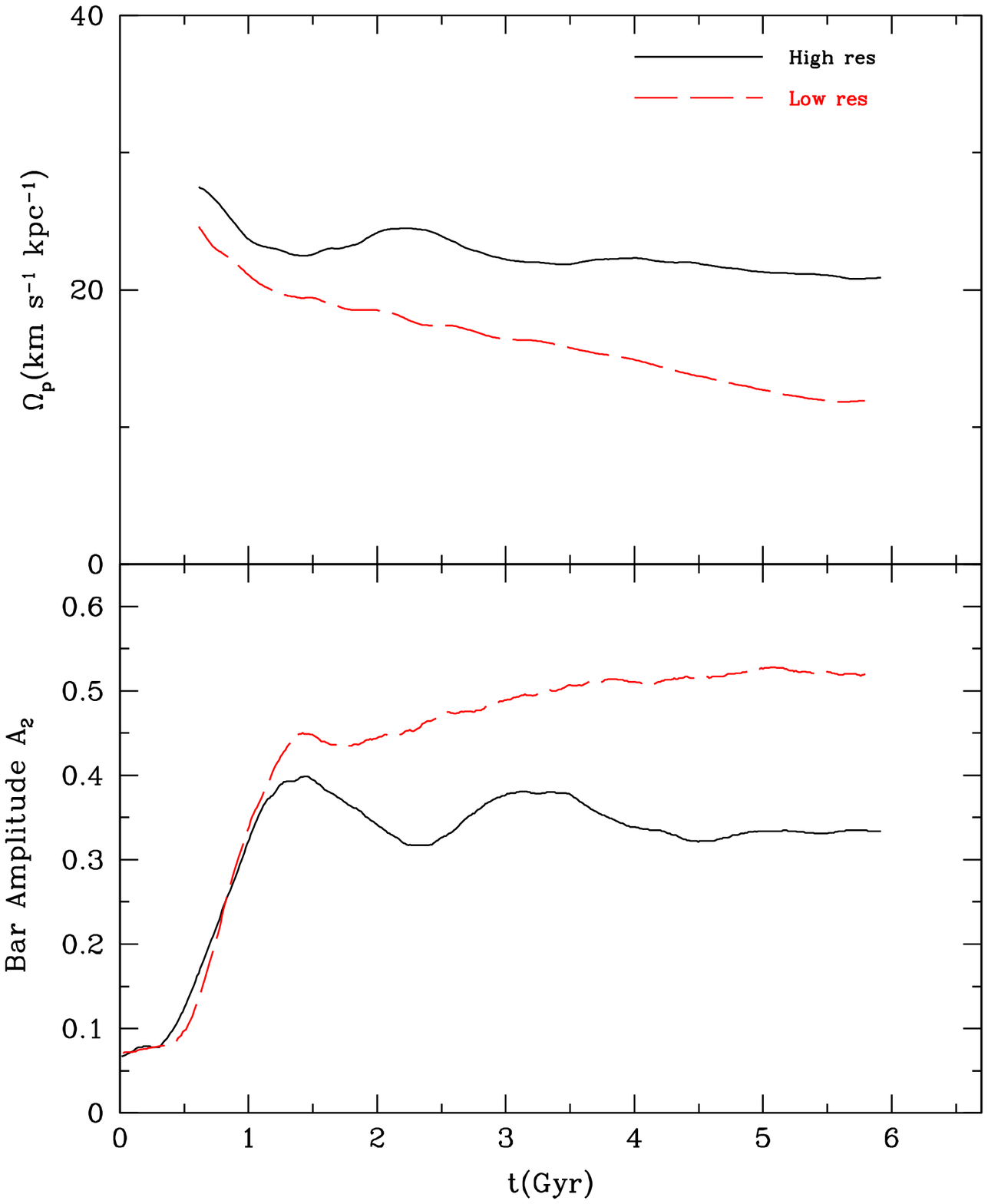}
\caption{
Dependence of bar properties on the force resolution. Evolution of the
pattern speed of the bar (top panel) and the bar amplitude (lower
panel) are shown for the high resolution model \kgtwo\ with force
resolution 112~pc)( full curves) and  for a low resolution simulation
with 560~pc resolution (dashed curves). The later simulation is the small time-step
model presented in Figure~\ref{fig:DTcon} with the full
curves. Increased force and mass resolution produce a more
concentrated bulge, which weakened the bar. A shorter and weaker bar
rotates faster and does not slow down much over many billions of
years.
\label{fig:ResCon}}
\end{figure}

Conditions for changing the time-step are different in different
codes. In the ART code the time-step decreases by factor 2  when the
number of particles exceeds some specified level (typically 2-4
particles).  A cell  that exceeds this level is split into eight
smaller cells resulting in the drop by $2^3$ times of the number of
particles in a cell. This prescription gives scaling of the time-step
with the local density $\rho$ as $dt \propto
\rho^{-1/3}$. \citet{cvk06} give more details of the procedure.  The
time-step in Quinn et al. scheme scales as $dt \propto
\rho^{-1/2}$.  \citet{Zemp07} also advocate a scheme with this scaling of the time-step.
The Gadget and Pkdgrav code use a scaling with the
gravitational acceleration $dt \propto g^{-1/2}$, which for
$\rho\propto r^{-2}$ gives $dt \propto \rho^{-1/4}$. Among all the
codes the Quinn et al. scheme uses the most aggressive prescription
for changing the time-step and Gadget has the smallest change in
$dt$. 

In practice, the time-step $dt$ for the Gadget-2 code is defined by
parameter 
\begin{equation}
\eta^2 =\frac{|{\bf g}|dt^2}{2\epsilon},
\end{equation}
\noindent  where $|{\bf g}|$ is the
acceleration, and $\epsilon$ is the effective Plummer softening. For
example, we used $\eta=0.04$ for \dgone~ and \kgtwo~ models.  It
is $\eta=0.11$ for \kgone~ model. In simulations where we increase
the time-step by factor two, we double $\eta$. For the Pkdgrav code
the time-step is defined by a similar expression: $\eta^2 =$ $|{\bf
  g}|dt^2/\epsilon_s$, where $\epsilon_s$ is the spline softening. We
use $\eta =0.025$ for model \ptwo~ and $\eta =0.050$ for model \pone.

 In all codes the time-step changes discretely by
factor two: $dt = dt_02^{-m}$, where $dt_0$ is the maximum time-step
and $m$ is an integer defined by local conditions (local density or
local acceleration).

\subsection{Numerical Effects: resolution and time-stepping}
\begin{figure*}
\includegraphics[width=0.8\textwidth]{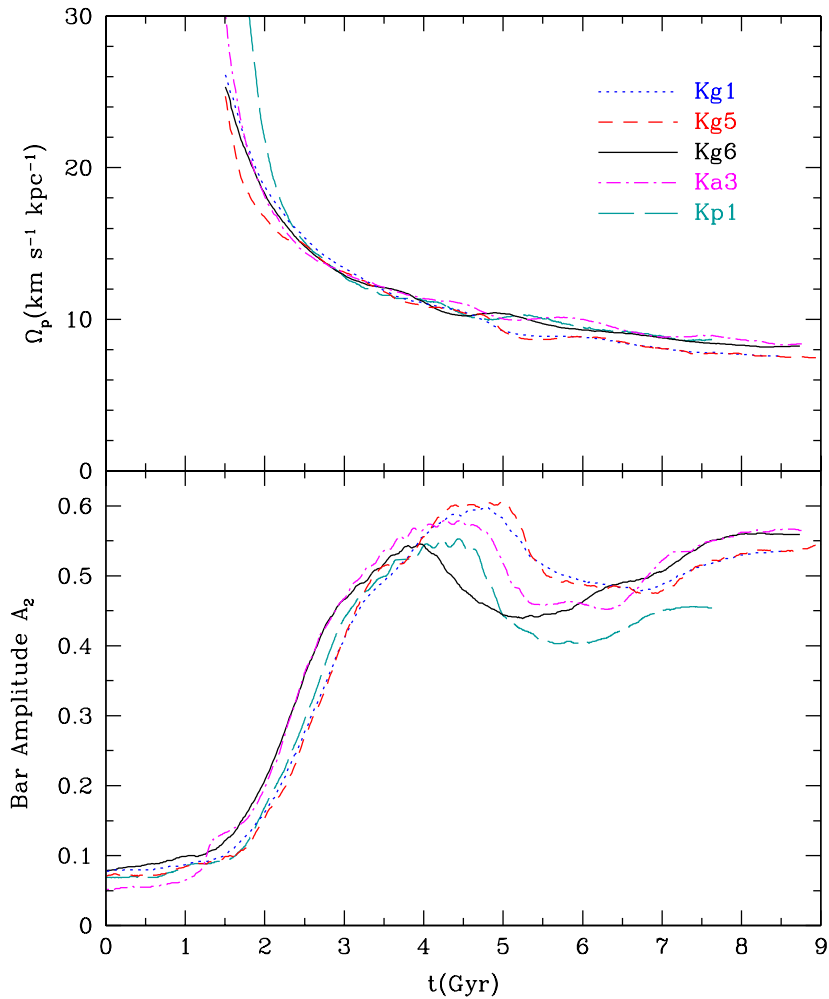}
\caption[fig:KhbTk]{
The evolution of the bar pattern speed $\Omega_p$ (top panel) and the
bar amplitude $A_2$ (lower panel) for K~ thick-disk models. All these
models evolve similarly: it takes $\sim 2-3$\,Gyrs to form a bar; the
buckling phase happens at $t\sim 5$\,Gyrs followed by  a
regime of a constant amplitude and nearly constant pattern
speed. Through the course of evolution the pattern speed declines by a
factor 2-3 and it is very low.
\label{fig:KhbTk}}
\end{figure*}
In order to obtain accurate results the trajectories of particles
should be integrated with a sufficient precision. There is no reliable
method of estimating how small a time-step should be. While there are
theoretical arguments what integration schemes should (or should not) be
used \citep{Quinn97, Preto99, volker05}, only tests can tell how
accurate results are.  Simulations of bars have a special reason why
the orbits should be accurate. Particles, which make up the bar move
typically on quite elongated trajectories periodically coming close to
the center. When this happens, the acceleration changes substantially,
and fast changing accelerations pose problems for numerical
integration. If accuracy of integration is not sufficient, a particle
may erroneously change its direction of motion and start moving away
from the bar. This artificial scattering on the center results in a
smaller number of particles staying in the bar.  Thus,
it is important to have not only accurate particle energies,
but also to have accurate phases of trajectories. The later is more
difficult than the former because different leap-frog schemes used in
current $N-$body codes are known to have problems with accurately
tracking orbital phases \citep[e.g.,][]{volker05}.  

The problem is more complicated because there is a real scattering on a
dense central region: trajectories have a tendency to get deflected,
when they come close to the center. The magnitude of this effect
depends on the mass and size of the central concentration.  Here the
force and mass resolution play important role. If the resolution is
not sufficient, the density in the central region will be lower
resulting in less deflection of orbits. Thus, more particles will stay
in the bar, which starts to trap even more particles. This leads to
excessive growth of the bar.


In order to estimate the effects of time integration of orbits, we
start by making simulations of the same model using different
time-steps.   We use the
Gadget-2 code to make simulations of the model K with low force
resolution of 560~pc and with $N_{disk} = 10^5$. The initial disk
height is $h_z=200$~pc.  Figure~\ref{fig:DTcon} shows results for
simulations with the time-step changing by factor two from one
simulation the other. Overall, we clearly see convergence of the
results, but it is not monotonic. Runs with different time-steps
evolve very similarly until (2.5--3)~Gyrs, when they start to
diverge. The reason for the divergence is likely to be related with
the fact that at around 2--3 Gyrs the system goes through the buckling
instability. At this stage the errors in orbit integration produce
large errors in the system configuration. For this particular model
the time-step of $10^5$~yrs was not sufficient: it resulted in a
qualitatively wrong answer.

It is difficult to predict what actually happens when the time-step is
not small enough. In this case it produced a steep decline in the bar
amplitude during the buckling stage. We had the same effect (weakened
bar) in ART simulations done with too large time-steps. At the same time,
it is quite possible to have just an opposite effect: an artificially stronger bar
with lower pattern speed. For example, \citet{sd06} run a model
similar to \kgtwo~  with a grossly insufficient time-step $1.5\times
10^5$~yrs; they got a very long and a slow bar. For the same model
\citet{vk03} used a ten times smaller step and got a significantly
shorter and faster bar.

The accuracy of orbit integration greatly depends on the distribution
of density in the central region: the steeper is the density profile,
the more difficult is the simulation. In our case the models K with a
thick disk ($h_z=714$~pc; e.g., runs \kgfive~ and \athree) did not form
a dense center. We find that for these models a relatively large step
of $dt\approx 10^5$~yrs is sufficient. Models K with  thin disk
($h_z=200$~pc) produce a nearly flat circular velocity curve implying a
steep profile, which can be roughly approximated as $\rho \propto
r^{-2}$ in the central 2~kpc region. These models showed inconsistent
results when  large time-steps $dt > 10^5$~yrs were used. Only when we
changed to much smaller steps ($dt \sim 10^4$~yrs), results became stable.

The observed effect of time-stepping presented in
Figure~\ref{fig:DTcon} is somewhat unexpected. The most difficult and
important part of the simulation is the motion of particles in the
central region. Taking a typical particle velocity of 200~km/s we
estimate that it would take $2\times 10^7$~yrs and 200 time-steps ($dt
=10^5$~yrs) for a particle to cross the central 2~kpc region. One
would expect that 200 time-steps is enough to provide reasonably
accurate results. Unfortunately, this is not the case as 
Figure~\ref{fig:DTcon} shows.  Indeed, our simple tests described
below indicate that in spite of the fact that the energy is reasonably
well preserved, the orbital angle is not --
trajectories are scattered very substantially when a large time-step
is used. This numerical scattering may result in incorrect properties
of the bar.

We investigate the situation with the accuracy of orbit integration by
studying the motion of particles in an idealized, but realistic case of
a spherical logarithmic potential $\phi(r) = \log(r)$.  We implemented four
time-integration schemes used in our simulations: a constant leapfrog
scheme, and block-schemes with a variable time-step used in ART,
Gadget-2/Pkdgrav, and Quinn et al. codes.

Figure~\ref{fig:Move1} shows the results of integration of an eccentric
orbit with the apocenter/pericenter ratio of about 10:1. The
gravitational potential is normalized in such a way that that the
binding energy is equal to the time averaged kinetic energy.  In order
to find the accuracy of the orbital angle, we run the orbit with a
very small time-step: $10^5$ steps per period. Then we find the
position angle of a low-accuracy run and compare it with the position
angle at the same moment in the small-step run.  As compared with a
constant-step run, all variable-step integration schemes give smaller
errors in the energy conservation, but the errors in the position
angle are too large.  The constant step run is definitely better, but
even in this case the errors ($\approx 10^o$) are still large.
Decreasing the time-step by factor of ten gives much better results as
shown by Figure~\ref{fig:Move2}. In this case even the phase of the
orbit is accurately simulated. The difference between the ART and the
Gadget codes is due to the fact that the ART code takes more small
steps in the central part of the orbit (the total number of steps is
the same). When we apply the Gadget time-step changing algorithm to
the ART code, we get exactly the same results. Quinn et al. takes even more
refinements in the center, which improves the accuracy. Yet, the code
gives more accurate results even when it runs with the same time-step
as Gadget. Still, for a 200 orbits integration -- typical for both the
bar simulations and for existing cosmological runs -- the differences
between the codes are not essential. The Quinn et al. scheme wins when
we make much longer runs: for 20,000 orbits with 200 step/orbit the
Quinn et al. code gave 10 times better accuracy than ART, and GADGET was
another factor of ten worse than ART.

Adequate force and mass resolution is another important numerical
issue, which is even more difficult to handle. The ``converged'' (in
the sense of time-stepping) solution presented in
Figure~\ref{fig:DTcon} has a strong bar, which dramatically slows
down: the pattern speed declines by a factor of two over 5~Gyr period.
Results presented in Figure~\ref{fig:ResCon} tell another story: a
low-resolution simulation can be very misleading. In this case we use
the same thin disk model with twice as many particles and run it with
five times better force resolution. The high mass and force resolution
simulation is qualitatively different: the bar is weaker and its
pattern speed hardly changes.

\section{Results}
\label{sec:Results}

Table~2 gives different properties of the simulated models as measured
after 5~Gyrs of evolution. In column 2 we present the fraction of the
angular momentum lost by the stellar material. The pattern speed
$\omp$ and the ratio of corotation radius to the bar radius are
presented in columns 3 and 4. The bar length is given in column 5. The
last three columns give parameters of a double-exponential
approximation of the stellar surface density: $\Sigma(r) =$
$\Sigma_{\rm bulge}\exp(-r/R_{\rm bulge})+ $ $\Sigma_{\rm
  disk}\exp(-r/R_{\rm disk})$.  

The models are clearly split into two
groups: those, which started with a thick disk (\athree, \kgone,
\kgfive, and \pone) and the models, which started with a thin disk.
For example, the ratio of the corotation radius to the bar radius is
about $1.7-1.8$ for thick disk models. For thin disk models the
ratio is visibly smaller: $1.2-1.4$. Thin-disk models have shorter
bars and less massive bulges. The differences are especially striking
when we compare simulations done with the same code and with similar
parameters. For example, models \atwo~ and \athree~ have very
similar numerical parameters (time-step, resolution, and number of
particles). Yet, their parameters at 5~Gyrs (and actually at any
moment) are drastically different. For example, the pattern speed for
model \atwo~ is 2.5 times larger than for the model \athree.

\begin{table*}
\caption{Parameters of the models after 5 Gyrs of evolution}
\begin{tabular}{lccccccc}
\hline
Name & $\Delta L/L$ & Pattern speed & $R_{\rm cor}/R_{\rm bar}$ & 
Bar length & Disk scale & Bulge scale & Bulge/Total \\
 &          & $\Omega_b$& & $R_b$ & length $R_d~$ &length~$R_{\rm bulge}$ & \\
 &  (\%) & ~(Gyr$^{-1}$)& & ~(kpc) & (kpc) & (kpc) & \\
(1) & (2) & (3) & (4) & (5) & (6) &(7) & (8)\\
\hline
& \multicolumn{7}{l}{Thick disk models} \\
 \athree    & 9.9  & 10.0  &  1.7  & 10.8  & 6.7  & 1.07 & 0.38\\
 \kgone    & 8.3  & 10.3  &  1.7  & 10.4  & 7.0   & 0.97 & 0.32\\
 \kgfive     & 10.3  & 9.1  & 1.7   & 12.1  & 6.6  & 1.10 &0.36\\
 \kgsix     & 8.2  & 10.5  & 1.5   & 11.5  & 6.5 & 1.03 &0.35\\
\pone      &  8.5& 10.3  &  1.8  &  10.0 & 6.4  & 1.00 & 0.33\\
 & \multicolumn{7}{l}{Thin disk models} \\
  \aone      & 5.1   & 21.0   & 1.25   & 6.7   & 5.7 & 0.73 & 0.26           \\
  \atwo       & 3.3   & 25.4   & 1.16   & 6.5    & 5.2  &0.54 & 0.20\\
 \kgtwo     & 2.7   & 19.8   &1.15    &7.5   &5.8   &0.63 & 0.23\\
 \kgthree & 2.6  & 21.7  & 1.20 &  6.7 & 5.3  &0.64 & 0.21 \\
 \kgfour    & 2.1    & 22.8  &1.22   & 6.0   &5.3   &0.61 & 0.21\\
 \ptwo      & 3.1    & 19.3  & 1.23  & 7.4   & 5.5  & 0.66& 0.22\\
\dgone      & 10.1  & 19.5  & 1.4   & 7.6  & 5.2  & 0.43 & 0.35\\
\hline
\end{tabular}
\end{table*}

\subsection{Thick Disk  Models}
\begin{figure}
\includegraphics[width=0.45\textwidth]{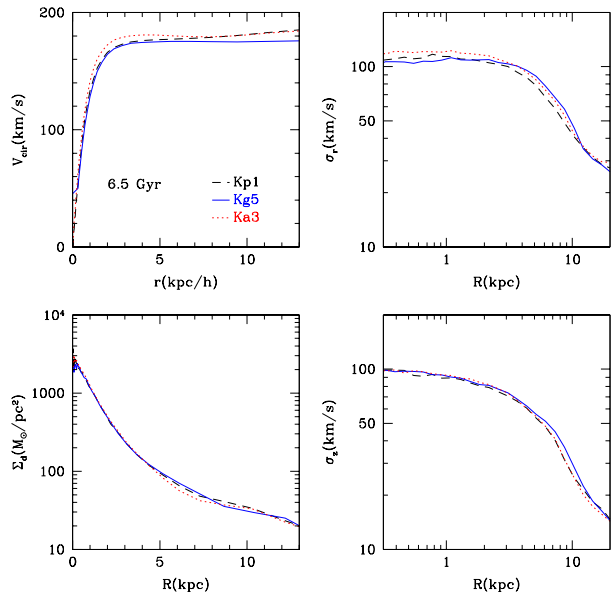}
\caption[fig:ProTk]{Profiles of total circular velocity (top-left panel),
stellar radial and vertical velocity dispersions (top- and bottom-right
panel, respectively), and surface density (bottom-left panel)
are shown in this four-panel plot for thick-disk K models
at 6.5 Gyr. Here we have selected one model for each code. 
They agree with each other remarkably well.
\label{fig:ProTk}}
\end{figure}

We start with analysis of thick disk K models.  
 It takes $\sim 2-3$\,Gyrs to form a bar; the
buckling phase happens at $t\sim 5$\,Gyrs followed by  a
regime of a constant amplitude and nearly constant pattern
speed. Figure~\ref{fig:KhbTk}
shows the evolution of the bar pattern speed \omp\ and the amplitude
of the bar $A_2$ for the models.  In Figure~\ref{fig:ProTk} we present
the total circular velocity profile $\sqrt{GM(<r)/r}$, the radial and
vertical stellar velocity dispersions, and the disk surface density as
a function of distance to the center of the galaxy, $R$. The
comparison is made at 6.5~Gyrs for three models: \athree, \kgfive, and
\pone. 

\begin{figure}
\includegraphics[width=0.475\textwidth]{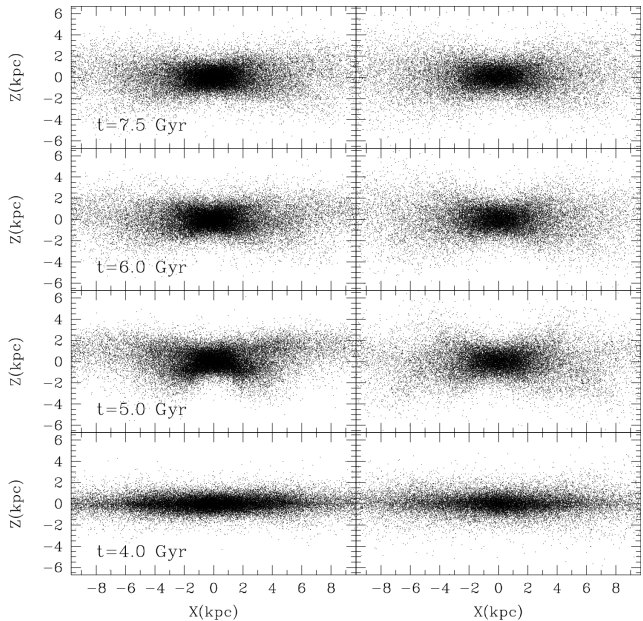}
\caption[fig:DisXZ]{Distribution of disk particles seen along the bar minor
axis at four different epochs: \kgfive\ model (left panels) and \athree\ model 
(right panels). Only particles with 
$|y| < 1.0\ \kpc$ are shown. In order to have similar number of 
particles in left and right panels only half of the particles, randomly 
chosen, were used in model \athree. At moment 5~Gyr a bar buckling  
is clearly seen.
\label{fig:DisXZ}}
\end{figure}

Figure~\ref{fig:DisXZ} shows disk particles seen along the minor axis
of the bar for K models with the thick disk \kgfive\ (left panels) and
\athree\ (right panels) at four different epochs. The selected time
moments represent different stages of bar evolution. At 4~Gyrs the bar
is the strongest, and it has not yet started the buckling stage: the
disk is still thin. At 5~Gyrs the system goes through the buckling
instability. At this moment the models look different. The
differences die out as the buckling instability proceeds.  Once the
buckling stage is finished, the models again are very close.

\begin{figure*}
\includegraphics[width=0.8\textwidth]{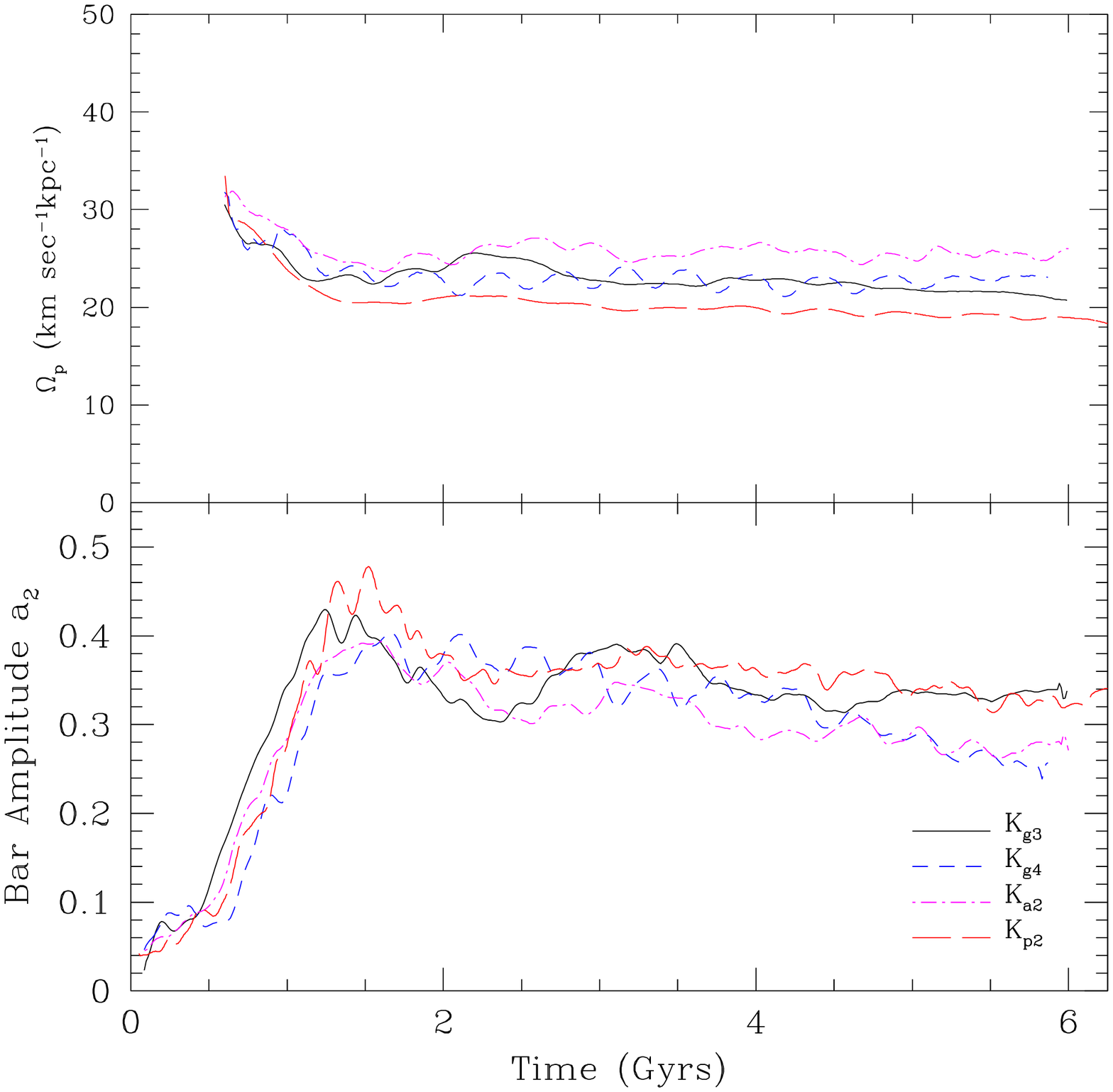}
\caption[fig:KhbTn]{
The evolution of the bar pattern speed (top panel) and of the bar
amplitude (lower panel) for thin-disk K models made with the Gadget-2,
the ART, and the Pkdgrav codes. All the models show pattern speeds
$\Omega_p$, which do not show a systematic decline with time.  There are differences
between different runs, but they are relatively small: $\Omega_p
=(21.6\pm 3) {\rm Gyr}^{-1}$.  Note substantial differences with
respect to the thick-disk models presented in Figure~\ref{fig:KhbTk}.
 \label{fig:KhbTn}}
\end{figure*}

Overall, the models evolve very similarly.  Some small differences are
observed during the buckling phase. Yet, models tend to converge at
the end of the evolution. The degree of agreement between different
codes as demonstrated by Figure~\ref{fig:ProTk} is remarkable. Inside
the central 5~kpc region the surface density of the disk deviates from
model to model by only few percent. The vertical velocity dispersion
over  the whole disk deviates not more than 5~km/s. 

This agreement between models demonstrates that the results are code
independent: if simulations are done with sufficiently small
time-steps and with similar force and mass resolution, all codes
produce nearly the same results. The results also show that there are
no problems with any particular code: all codes produce the same
``answer''.

\subsection{Thin Disk Models}

\begin{figure}
\includegraphics[width=0.45\textwidth]{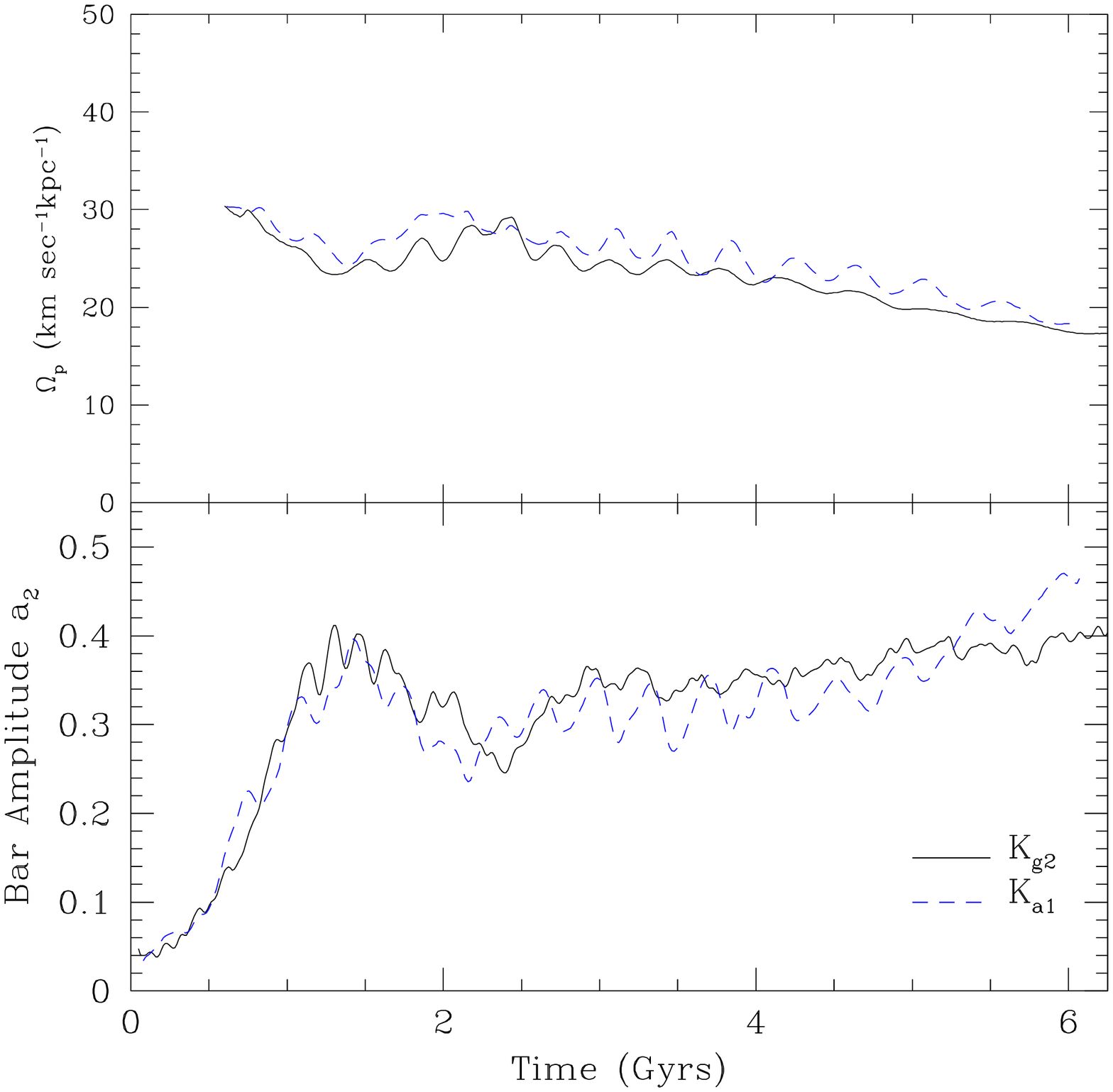}
\caption[fig:KTn]{
The evolution of the bar pattern speed (top panel) and of the bar
amplitude (lower panel) for  thin-disk models \kgtwo~ (Gadget-2) and \aone~ (ART).
These two simulations were performed
with different codes, yet they  produce very similar results. 
\label{fig:KTn}}
\end{figure}

Figures~\ref{fig:KhbTn} and ~\ref{fig:KTn} show the evolution of the
bar amplitude and the bar pattern speed for thin disk K models. Again,
the simulations behave very similarly, but this time they are very
different from the thick disk models K.  The thin disk models do not
show any substantial evolution in the bar pattern speed and in the bar
amplitude.  Nevertheless, the agreement between the simulations is not
as close as in the case of the thick disk models. It is important to
note that there are no systematic differences between results obtained
with different codes.  The comparison of two runs presented in
Figure~\ref{fig:KTn} is especially striking.  For example, the Gadget
model \kgtwo~ shows some small decline in the pattern speed and some
oscillations in bar pattern speed and bar amplitude. The same
evolution of $\Omega_p$ is demonstrated by the ART model \aone, which
also has the oscillations with a slightly larger amplitude.  The bar
amplitudes for the two models are also very close over the whole
simulated period of time. At the same time, models \atwo~ (ART) and
\kgfour~ (Gadget), which give very similar results, do not show any
indication of slowing down of the bar (see Figure~\ref{fig:KhbTn}).

We believe that this agreement between results produced by different
codes indicates that there are real physical differences between the
models and that those differences are not of numerical origin. In
other words, when started with the same physical initial conditions,
the evolution produces different results. In the case of the thin disk
K models the differences are on the level of $\sim 10$ percent after
6~Gyrs of evolution.

What are the possible reasons for these variations between the models?
The $N$-body problem is deterministic: initial conditions
uniquely define the outcome of evolution.  If one starts with
identical initial conditions and uses a perfect code, the results must
be unique. Yet, this is too simplistic. An unstable system may have
divergent evolutionary tracks. A fluctuation slightly changes the
system, but because of instabilities, the fluctuation grows, and the
final answer is different as compared with the evolution of the system
without the fluctuation.  The barred stellar dynamical models have two
stages when a system is unstable: the initial stage of formation of
the bar and the buckling instability. In addition, even in the quiet
periods, the bar itself is an example of a potentially unstable
system. For example, it could start trapping more particles resulting
in even stronger bar, which traps even more particles. To large
degree, this is exactly what happened with the bars in thick disk
K models. The bars were expanding until all the disk became the bar.

The source of fluctuations is an interesting issue. In the simulations
the fluctuations are related with the initial random phases and
amplitudes and with numerical inaccuracies. Yet, we should not forget
that our models are only approximations to the reality. In real
galaxies there is no shortage of fluctuations including satellite
galaxies and molecular clouds to name the few. Whatever is the source
of fluctuations, simulations of the same physical model evolve
slightly different and produce slightly different results. This is
exactly what the thin disk models did.

\begin{figure}
\includegraphics[width=0.47\textwidth]{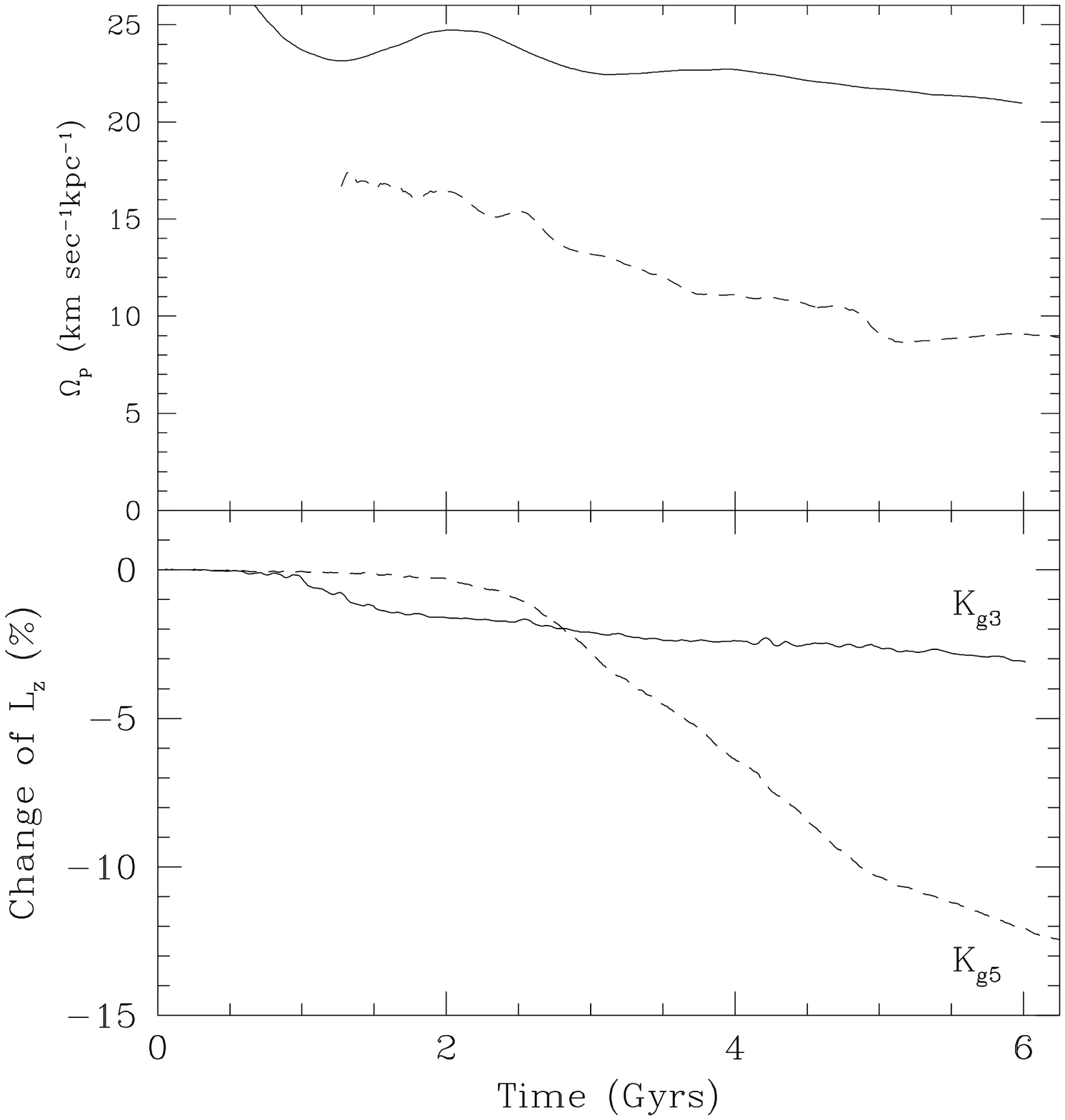}
\caption[fig:Lzkhb]{
Effects of the initial phase-space density: the evolution of the
pattern speed (top panel) and the disk angular momentum loss (bottom
panel) for model \kgthree\ (full curves) and model \kgfive (dashed
curves). These two models differ by the initial value of the scale
height of the disk, which results in substantially different
phase-space density. The thin model goes through the buckling
instability earlier and has a smaller angular momentum loss.
\label{fig:Lzkhb}}
\end{figure}

\subsection{Effects of the phase-space density: thin versus thick
  disks}

The only difference in the initial conditions between the thin disk
and thick disk K models is the disk scale height. All the other parameters are
the same.  These seemingly minor variations in initial conditions
resulted in a remarkably different evolution and in large differences
in the structure of the evolved models.  In order to highlight the
differences, Figure \ref{fig:Lzkhb} shows the angular momentum loss
$\Delta L/L$ of the stellar disk and the bar pattern speed for two
models run with the Gadget code: one with a thin disk (\kgthree) and
another with a thick disk (\kgfive). The disk in the thick disk model
loses four times more angular momentum and its bar rotates 2.5
times slower as compared with the thin disk model. The dark matter is
the same in both models, and it cannot be the reason of the
difference.

So, what are the possible reasons for such a large effect of the disk
height?  The thickness of the stellar disk affects the vertical waves
and oscillations \citep{Merritt1994}. This effect is definitely
present and will have some impact on the evolution.  At the same time,
those vertical modes are likely to play a significant role during the
buckling stage \citep{Merritt1994}.  However, the differences between
the models develop too early and they are too large for the vertical
modes to be the culprit.   In the absence of a reliable theory of
stellar bars, one can only speculate what is going on.  One may think
about two other effects, which can influence the systems: the Jeans
mass and the phase-space density. The random velocities, which define
the Jeans mass, tend to prevent collapse of perturbations on small
scales. Thus, for the same random velocities, larger Jeans
masses imply smaller densities.  The phase-space density acts in the
same way.  Thus, we expect that models with higher phase-space density
(or small Jeans mass) will result in denser and more compact central
region, which affects the growth of bars in a profound way.

For a fixed surface density the vertical disk height $h_z$ changes the
density in the disk $\rho\propto h_z^{-1}$ and the velocity dispersion
$\sigma_z^2 \propto h_z$ (see eq.(\ref{eq:height})). Assuming for simplicity that the Jeans mass
scales with the velocities and density as $M_J\propto
\sigma_R\sigma_\phi\sigma_z/\sqrt\rho$, we get $M_J\propto h_z$, which
is a remarkably strong effect considering that in our models $h_z$
varies by factor 3.5. The phase-space density changes even more:
$f=\rho/ \sigma_R\sigma_\phi\sigma_z \propto h_z^{-3/2}$. For our
thin/thick models this gives variations by factor 7. 

The Jeans mass affects the evolution in a similar way as the Toomre
stability parameter $Q$: larger $Q$ results in later formation of the
bar and in less prominent spiral arms. Comparison of the initial
stages of bar formation in Figures~\ref{fig:KhbTk} and \ref{fig:KhbTn}
shows the  effect: larger $h_z$ (thus, larger $\sigma_z$) results
in much delayed formation of the bar.

\begin{figure}
\includegraphics[width=0.45\textwidth]{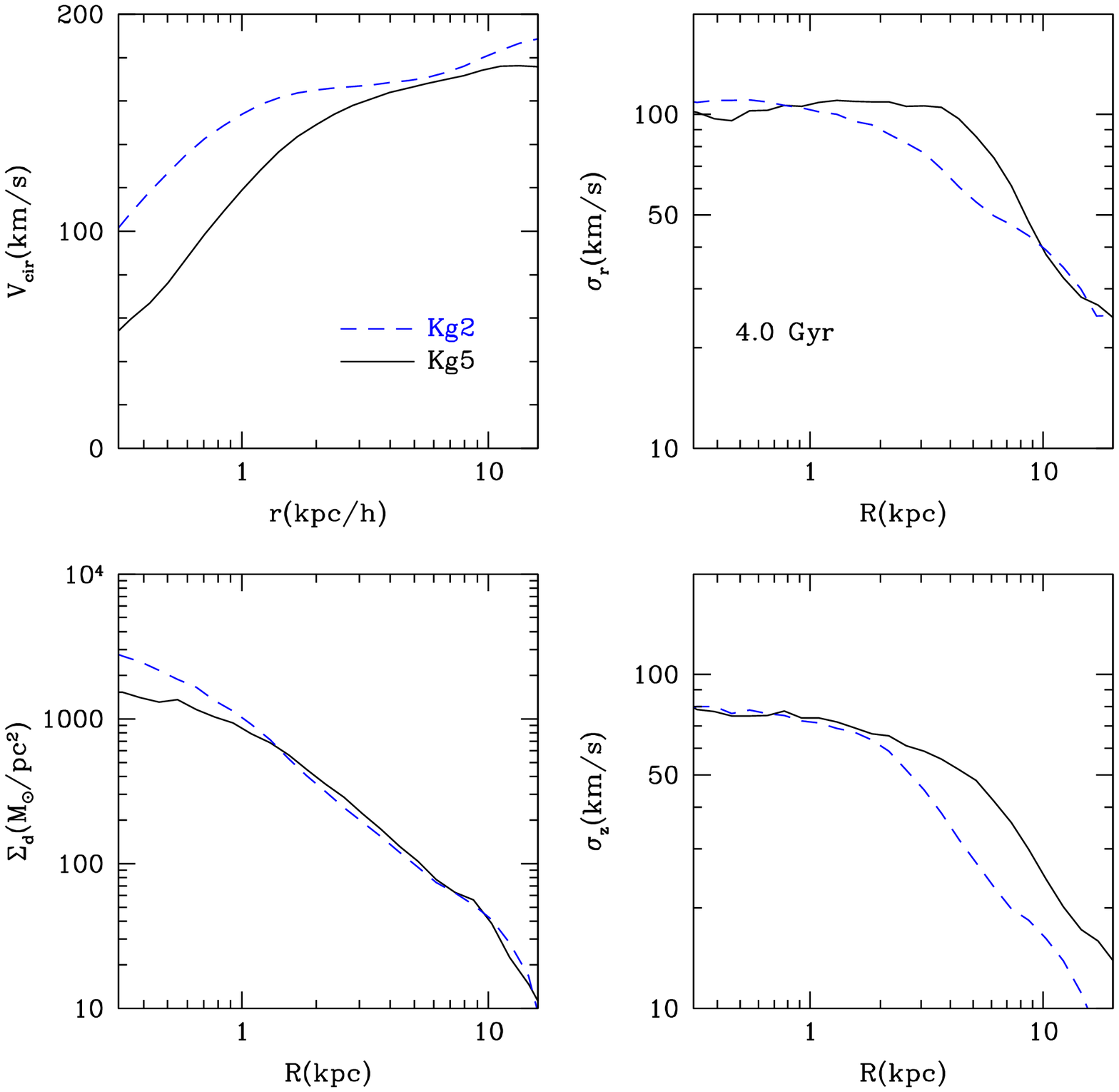}
\caption[fig:TnTk]{
Comparison of profiles of models with different initial phase-space
density. The circular velocity (top-left panel), the stellar radial
and vertical velocity dispersions (top- and bottom-right panel,
respectively), and the surface density (bottom-left panel), as a
function of radius for models \kgtwo\ and \kgfive, thin and thick
disk, respectively.  The comparison is made at 4~Gyr.  The thin disk
model (dashed curves) evolves to a more concentrated structure.  The
central concentration explains why the bar pattern speed for this
model is higher than in the thick disk model.

\label{fig:TnTk}}
\end{figure}

\begin{table*}
\caption{Comparison of Parameters of the Milky Way  and the model \kgthree}
\begin{tabular}{lccl}
\hline
Parameter& \kgthree & Milky Way & Reference\\
\hline
Circular velocity (km/s)    &  220 &  210-230 & \citet[][Sec.10.6]{Binney98}\\
Surface disk density at $R_\odot$ ($M_\odot /pc^{2}$) &   44.6      &    $48\pm 9$     &   \citet{Kuijken91}     \\
Vertical rms velocity of stars at $R_\odot$ (km/s) &14 & 15-20 & \citet{Dehnen98b}\\
Radial rms velocity of stars at $R_\odot$ (km/s) & 38&35-40 &  \citet{Dehnen98b}\\
Pattern Speed $\Omega_p$ (km/s/kpc)& 50& $53\pm 3$ & \citet{Dehnen99} \\
Bar length (kpc) &3.3 & 3.0-3.5&\citet{Freudenreich98} \\
Total mass inside 60~kpc ($10^{11}M_\odot$)& 5.5& $4\pm 0.7$& \citet{Xue08} \\
Total mass inside 100~kpc ($10^{11}M_\odot$)& 7.3& $7\pm 2.5$& \citet{Dehnen98} \\
\end{tabular}
\label{tab:MW}
\end{table*}

\begin{figure}
\includegraphics[width=0.45\textwidth]{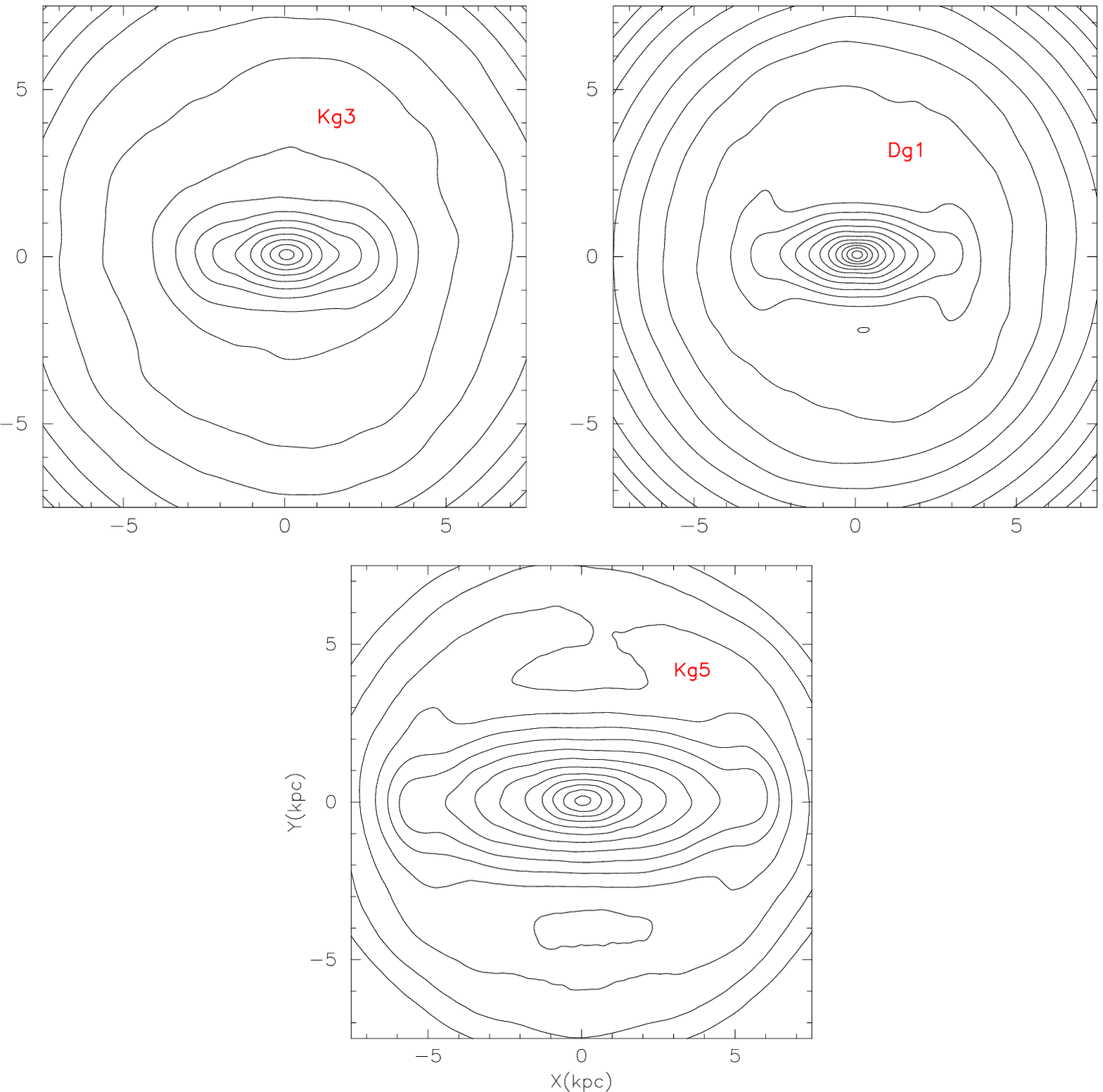}
\caption[fig:Surface]{ The iso-contours of the surface stellar density
  for representative models.  The thin disk models \kgthree~ and
  \dgone~ develop realistic bars with the ratio of corotation radius
  to the bar length ${\cal R} =1.2$ (left panel; $R_{\rm
    cor}=4.5$~kpc) and ${\cal R} =1.4$ (right panel; $R_{\rm
    cor}=6.1$~kpc). The thick disk model \kgfive~ (bottom panel) has a
  bar, that covers the whole disk with the corotation radius
  9.3~kpc.  Distances in the plot are given in kpc units. All the
  models are re-scaled to have the disk scale length 3~kpc -- the same
  as for the Milky Way galaxy.}
\label{fig:Surface}
\end{figure}

The phase-space density is a complicated quantity, which in practice
is used in the form of a coarse-grained phase-space density.
\citet{Avila2005} present detailed results on the evolution of the
phase-space density during the formation and evolution of barred disk
models. As the system evolves, the coarse-grained phase-space density
$f= \rho/ \sigma_R\sigma_\phi\sigma_z$ decreases.  The phase-space
density $f$ can be considered as a measure of the degree of
compressibility of a gravitational system: for given rms velocities
it defines the real-space density: $\rho = f
\sigma_R\sigma_\phi\sigma_z$. The larger is the phase-space density,
the larger is $\rho$. In turn, the density in the central region of
a bar is an important factor because according to our results it moderates the growth of the
bar and can prevent it from growing excessively. Thus, the
phase-space density in the central disk region is a fundamental
parameter, which significantly affects the evolution of barred
galaxies.

Indeed, the buildup of mass in the high phase-space density models
happens in the models.  Figure \ref{fig:TnTk} shows the total (stellar
plus dark matter) circular velocity (top-panel) and the stellar surface
density (bottom-panel) profiles after 4 Gyr of evolution. We see that
the model \kgtwo\ (thin disk) has a larger inner circular velocity and
a larger surface density as compared with the low phase-space density
model \kgfive. The rms velocities (right panels) are nearly the same
for the two models. Thus, the \kgtwo\ model has a substantially larger
(by a factor of four) phase-space density. It started with a larger $f$
and it ended with a larger $f$.

\section{Discussion and conclusions}
\label{sec:Discussion}

\begin{figure}
\includegraphics[width=0.475\textwidth]{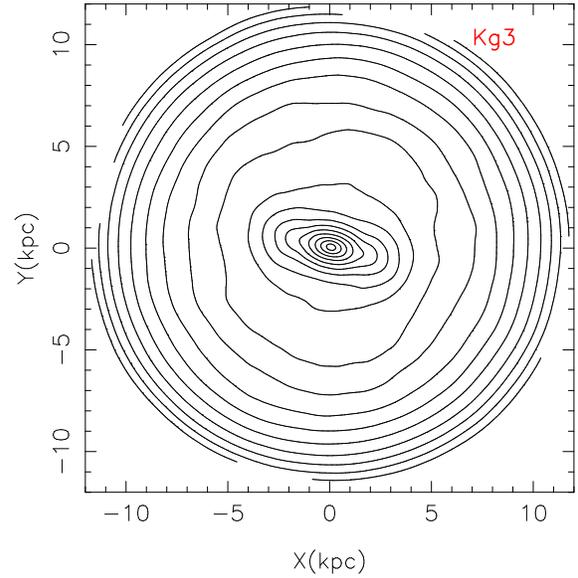}
\includegraphics[width=0.465\textwidth]{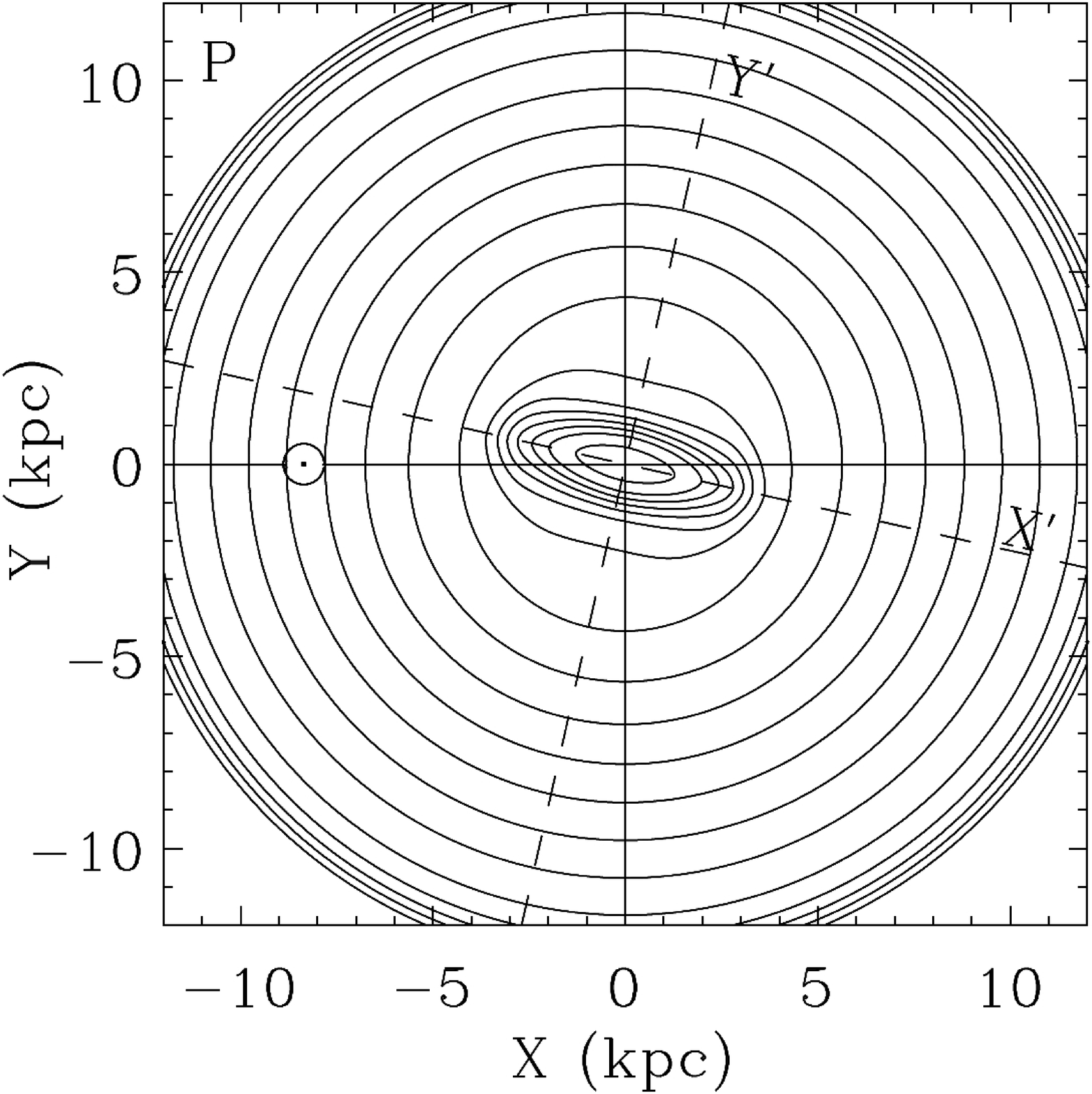}
\caption[fig:Surface]{The iso-contours of the surface stellar density
  for model \kgthree~ (top panel) and the Milky Way galaxy (bottom
  panel).  The \kgthree~ model was re-scaled to have the evolved disk
  scale length  3~kpc, which is close to the scale length of our
  Galaxy. The bottom panel shows one of the models from
  \citet[][fig. 14, right panel]{Freudenreich98}. The model represents
  a multi parameter fit to the COBE DIRBE maps of the near-infrared light
  coming from central regions of our Galaxy. The small circle shows
  position of the Sun.  The \kgthree~ model reproduces the length and
  the flattening of the Milky Way bar.}
\label{fig:SurfaceMW}
\end{figure}

We find that numerical effects -- the mass and force resolution, and
the time integration of trajectories -- can significantly alter
results of simulations. The time-step of integration must be small
enough to allow accurate integration of expected
trajectories. Accuracy of energy conservation can give a misleading
impression that the simulation is adequate, while it actually makes
substantial errors in position of orbits.  Our experiments with
realistic orbits in models with flat rotation curves indicate that
even the best available integration schemes require $\sim 2000$
time-steps per orbit. This requirement is valid for variable-step
schemes with more steps required for a constant-step
schemes. Numerical tests with full-scale dynamical models confirm the
condition.  This condition results in very small time-steps of $dt
\sim 10^4$~yrs. If one uses dimensionless units defined by scaling
$G=1$, $M_{\rm disk}=1$, $R_d =1$, then the dimensionless time step
should be $\tilde dt\approx 5\times 10^{-4}$. This should be compared
with typically used $\tilde dt \approx 0.01$
\citep[e.g.,][]{lia02}. This time-step would be insufficient for
integration of models with central mass concentration presented in
this paper \footnote{Models studied by \citet{lia02} are less
  concentrated and do not require a small time-step.}.  A constant
time-step $dt =5\times 10^5$~yrs used by \citet{Widrow08} is too large
for models of the Milky Way galaxy studied in their paper.

The mass and related with it the 
force resolution also play important role. As
Figure~\ref{fig:ResCon} shows, a low force resolution produces a less
dense central region, which results in a long and a very massive
bar. The same model run with better force resolution produces a
shorter, less massive, and faster rotating bar.

Once the necessary numerical conditions are fulfilled, the codes produce
practically the same results.  We do not find any systematic
deviations between the results obtained with TREE codes (Gadget and
Pkdrgav) and with the Adaptive-Mesh-Refinement (ART) code.

Disk height is an important parameter, which is often ignored in
models of barred galaxies. In the models, which we consider in this
paper, the disk height determines the global properties of the
bars. Figure~\ref{fig:Surface} shows surface density maps of the models
with different initial disk thickness. Models with thin disks
produce short bars with $R_{\rm bar}\approx R_d$, which rotate
relatively fast: ${\cal R} =1.2-1.4$ and which show very little
decline of the pattern speed. Models with  thick disk produce 
long and slow rotating bar.  In order to facilitate the comparison
with the our Galaxy, we re-scaled models to have the evolved disk
scale length 2.65~kpc and to have the circular velocity at the solar distance
220~km/s. Any $N$-body system has two arbitrary scaling factors, which
can be used to scale the system.

Having scaled the models to fit the disk scale length, we can compare
other parameters of the models. Because the simulations with thin disks
produce reasonable models, we use one of the models (\kgthree) and
compare it with the Milky Way. Table~\ref{tab:MW} gives a list of some
parameters.  Figure~\ref{fig:SurfaceMW} compares the surface density
maps of the Milky Way \citep{Freudenreich98} and the \kgthree~
model. These comparisons show that the model fits the Milky Way
reasonably well.

We suggest  that the disk height is
only an indicator of a more
fundamental property -- the phase-space density in the central
($R<R_d$) region. In our models the phase-space density is uniquely
related with the disk height.  Initially in our models the disk height
is low and the stability parameter $Q=1.3-1.8$ is constant across the
disk. Thus, the phase-space density in the central region is high, and
subsequent evolution brings that highly compressible stellar fluid
close to the center, where it forms a nearly flat circular velocity
curve. The later, as we speculate, is responsible for arresting the
growth of the bar.  This relation between the disk height and the
central phase-space density may not be true in general case. For
example, \citet{od03} and \citet{Widrow08} consider models with large
central $Q$ and, thus, with a low central phase-space density. Indeed,
in their models bars slow down substantially.

The nearly constant pattern speed of bars in the thin disk models is
somewhat puzzling. A bar is a very massive non-axisymmetric object,
which rotates inside a non-rotating dark matter halo. As such, one
might expect that it should experience dynamical friction and slow
down. This is why results of \citet{vk03}, which showed models with
little slowing down of bars, were met with skepticism. Simulations
presented in this paper confirm the findings of \citet{vk03} and show
that they cannot be related with numerical problems.

Orbital resonances may be responsible for the observed
slow dynamical friction. The resonances in barred galaxies have been
extensively studied in recent years using $N$-body simulations
\citep[e.g.,][]{lia03,cvk06,Ceverino07,Weinberg07}. It is now well
established that a large fraction of {\it stellar particles} are in
resonance with the bar. Some fraction of {\it dark matter particles} are
also in resonance \citep{cvk06,Ceverino07}, and these are very
important for the dynamical friction between the stellar bar and the
dark matter.

There are two types of (exact) resonances in an autonomous Hamiltonian
dynamical system: elliptic and hyperbolic \citep{Arnold}. Orbits,
which are close to an elliptic resonance librate around the exact
resonance and have the structure of a simple pendulum
\citep[][Sec.2.4]{Lich83}, \citep[][Sec. 8]{Murray99}. Hyperbolic
resonances are points on intersections of separatrixes, which separate
domains of elliptical resonances. Orbits close to a
hyperbolic resonance are unstable and tend to migrate away from the
resonance, while orbits close to an elliptical resonance are stable.

\citet{Ceverino07} study in detail the resonant orbits
 in simulations of barred galaxies. It appears that the orbits
belong to elliptical resonances. These orbits track their resonance:
the orbits do not evolve if the resonance does not move, and they
follow the resonances if it gradually migrates in the
phase space. Thus, the elliptical resonances trap the orbits: if an
orbit for whatever reason happens to appear in the domain of the
resonance, it will have a tendency to stay  with the
resonance. This phenomenon is thought to be common in Solar System
dynamics \citep[e.g.,][]{Malhotra93}.  Resonance trapping explains why the
simulations show maxima in the distribution of orbital frequencies at
the positions of resonances. 

\citet{cvk06} investigate another aspect of the resonant interaction
between the stellar bar and the dark matter. They found that the dark
matter particles, which are in resonance with the stellar bar,
themselves form a bar, which rotates with the same angular speed and
has a very small lag angle ($\sim 10^o$) as compared with the stellar
bar. \citet{cvk06} argue that the interaction between the dark matter
and the stellar bars is the main mechanism for the dynamical friction
between the disk and the dark matter. The near alignment of the dark
matter and stellar bars means that their interaction is minimized by
the resonances.

These results indicate that the resonant interaction between the
stellar bar and the dark matter is mostly due to elliptical
resonances, and, thus, has a tendency to {\it minimize} the transfer
of the angular momentum from the disk to the dark matter.  Following
orbits in such resonances emphasizes the need for conservative
time-steps in $N$-body simulations.  Also subtle changes in the
underlying global potential could change the relative number of orbits
in these resonances leading to disparate results for the slowing of
the bar.

\section*{Acknowledgments}

P.C. acknowledges support from the DGAPA-UNAM grant IN112806.  A.K.
acknowledges support by NSF grants to NMSU. T.Q. acknowledges support
by NSF ITR grant PHY-0205413.  O.V. thanks support from a CONACyT
Repatriacion Fellowship.  Some of the simulations presented in this
paper were performed on the HP CP 4000 cluster (Kan-Balam) at
DGSCA-UNAM, on the Columbia computer at the NASA Ames center, and at
the Artic Region Supercomputing Center.

\label{lastpage}

\end{document}